\documentclass[aps,prd,amsmath,twocolumn,longbibliography]{revtex4-1}

\usepackage{graphicx,amsmath,latexsym} 
\usepackage{subcaption}

\def\be{\begin{equation}}
\def\ee{\end{equation}}
\def\tz{{\tilde z}}

\newcommand{\eq}[1]{Eq.~(\ref{#1})}

\newcommand{\bfzeta}{\mbox{\boldmath$\zeta$}}
\def\bea{\begin{eqnarray}}
\def\eea{\end{eqnarray}}\newcommand{\bfk}{{\bf k}}
\def\s{\sigma}

\def\la{\langle}\def\ra{\rangle}\def\k{\kappa}\def\b{\beta}
\def\bfb{{\bf b}}
\def\r{\rho}

\def\D{\Delta}
\def\e{\epsilon}
\def\para{\parallel}
\begin{document}

\title{NT @ UW-21-13\\Confinement in Two-Dimensional QCD and the Infinitely Long Pion}

\author{Colin M. Weller and Gerald A. Miller}
 
\affiliation{
University of  Washington, Seattle  \\
Seattle, WA 98195-1560}

\date{\today}

\begin{abstract}
Three current models of QCD in (1+1) dimensions are examined and extended in light-front coordinates. A pion of high momentum is found to have an infinite extent along its direction of motion.  
\end{abstract}

\maketitle

 There  has been a recent revival of interest in theories involving two-dimensional(one-space-one time) treatments of QCD~\cite{DeTeramond:2021jnn,Li:2021jqb,Ahmady:2021lsh,Ahmady:2021yzh}.
This stems  from the need to include the effects of non-vanishing quark masses and also   to enlarge the number of space-time variables of light-front holographic QCD  to four.   In the original treatments (see the review~\cite{Brodsky:2014yha}) the chiral limit is used and the longitudinal light-front momentum fraction, $x$, is frozen~\cite{Sheckler:2020fbt}, so that effectively the only degrees of freedom are light-front time and transverse position.  The first effort aimed at including the effects of mass was contained in Ref.~\cite{Chabysheva:2012fe}. The present epistle aims to unify the earlier treatments and to exhibit the confining aspects of the approaches in three-spatial dimensions using the spatial coordinate, $\tz$, that is canonically conjugate to the variable $x$~\cite{Miller:2019ysh}.

%The resulting formalism is denoted  
%ight-front holographic QCD~\cite{Brodsky:2006uqa} in which the longitudinal degree of freedom, represented by the light-front momentum fraction, $x$. 

We begin by briefly summarizing light-front holographic QCD (LFHQCD) following \cite{Brodsky:2014yha}. Light-front  (LF) quantization is a relativistic, frame-independent approach to describing the constituent structure of hadrons. The simple structure of the light-front vacuum allows an unambiguous definition of the partonic content of a hadron in QCD and of hadronic light-front wave functions.
 %, the underlying link between large distance hadronic states and the constituent degrees of freedom at short distances.
 The QCD light-front Hamiltonian  is constructed from the energy-momentum tensor of QCD.   The spectrum and light-front wave functions of relativistic bound states are obtained from the resulting eigenvalue equation, an infinite set of coupled integral equations for the LF components in a complete basis of noninteracting $n$-particle states. This provides a quantum-mechanical probabilistic interpretation of the structure of hadronic states in terms of their constituents at the same light-front time $x^+=x^0+x^3$. 

To a first semiclassical approximation, where quantum loops and quark masses are not included, the relativistic bound-state equation for light hadrons can be reduced to an effective LF Schrödinger equation. In  conjugate coordinate  space, the relevant dynamical variable is an invariant impact kinematical variable $\bfzeta=\bfb\sqrt{x(1-x)}$, where $\bfb$ is the transverse separation of the constituents.  
(Boldface notation  specifies vectors of the two-dimensional transverse space.)
The  complexities of the strong interaction dynamics are then hidden in an effective potential $U(\zeta)$.  
Remarkably, the resulting light-front Hamiltonian has a structure  that matches exactly the eigenvalue equations in anti-deSitter space~\cite{Brodsky:2014yha}. This offers the possibility to explicitly connect the AdS wave function $\Phi(z)$ to the internal constituent structure of hadrons. Moreover, one can obtain the AdS wave equations by starting from the semiclassical approximation to light-front QCD in physical spacetime. This connection yields a relation between the coordinate $z$ of AdS space with the impact LF variable $\zeta$, thus giving the holographic variable $z$ a precise definition and intuitive meaning in light-front QCD~\cite{Brodsky:2009bd}.

 %%%%%%%%%%%%%%%%%%%%%%%%%%%%%%%%%%%%%%%%%%%%%%

An effective
light-front Schr\"odinger equation for the quark ($m_1)$-antiquark ($m_2)$,
wave function $\psi(x,\bfk),$ of a meson is given by
\be \label{eq:fullLFSE}
\left[ \frac{m_1^2}{x}+\frac{m_2^2}{1-x}+\frac{\bfk^2}{x(1-x)}
    + V_{\rm eff}\right]\psi=M_h^2\psi,
\ee
where $\bfk$ is the transverse relative momentum,   the first three terms are the light-front energy of two non-interacting partons, $M_h$ is the invariant mass of the hadron, and 
$ V_{\rm eff}$ is an effective potential that acts in three-dimensional space. 

Note that the kinetic energy term depends only on the two-dimensional vector $\bfk/\sqrt{x(1-x)}$, if the chiral limit ($m_{1,2}=0)$ is taken.
Then, using coordinate space, and taking $V_{\rm eff}$ to depend on $\bfzeta,\,V_{\rm eff}=U_\perp(\bfzeta)$ one obtains
\bea \left(-\frac{d^2}{d\zeta^2} + \frac{L^2}{\zeta^2}+U_\perp(\bfzeta)\right) \varphi(\bfzeta) = M^2 \varphi(\bfzeta),\label{2D}\eea
which is the same as the equation of motion in the soft-wall  AdS model~\cite{Karch:2006pv} if $L$ is taken to be angular momentum, $\zeta$ is identified with the fifth dimension $z$ in AdS space, and the wave function $\varphi(\bfzeta)$ is identified with  the string modes $\Phi(z)$.

Assuming  duality with AdS$_5$  suggests models for $U_{\perp}(\bfzeta)$, through
a correspondence between the transverse Schr\"odinger equation (\ref{eq:fullLFSE})
and the equation of motion for a spin-$J$ field in AdS$_5$~\cite{Brodsky:2014yha}.
 For the soft-wall model~\cite{Karch:2006pv},  the effective potential reduces to an oscillator potential
$
U_{\perp}(\zeta)=\kappa^4\zeta^2+2\kappa^2(J-1),
$
where $\k$ is the strength of the holographic confinement and $J$ is the total angular momentum.
For this potential, the spectrum of masses is $M_{\perp}^2=4\kappa^2\left(n+(J+|L_z|)/2\right)$,
with $n$ the radial quantum number, and the transverse wave functions are the
two-dimensional oscillator eigenfunctions.  The spectrum of the model provides
for a linear Regge trajectory and a good fit to light meson masses.
Using this potential leads to a massless pion ($n=0,L_z=0,J=0$) in the chiral limit.

In \eq{2D} the variable $x$ is held fixed, leading to a $(2+1)$ space-time description that  provides an excellent representation of the hadronic spectrum, but is manifestly incomplete because 
of the missing dynamics of the longitudinal direction. 
One incorporates \cite{DeTeramond:2021jnn,Li:2021jqb,Ahmady:2021lsh,Ahmady:2021yzh,Chabysheva:2012fe} these dynamics by asserting that the total potential $V_{\rm eff} $ of \eq{eq:fullLFSE} is the sum of two terms: $V_{\rm eff}=U_{\perp}(\bfzeta) +V_\para(x)$, with
\bea
\left[ \frac{m_1^2}{x}+\frac{m_2^2}{1-x} 
    + V_\para\right]X_n(x)=M_\para^2X_n(x).
\eea
Under these assumptions the full wave function $\psi$ is given by the product
\bea \psi(x,\bfb)=\varphi(\bfzeta)X_n(x)\eea
and $M_h^2=M_\perp^2+M_\para^2$. Here the normalization convention~\cite{Callan:1975ps}
\bea\int_0^1\frac{|X_n(x)|^2}{x(1-x)}\,dx=1\label{norm}\eea 
is used. It is helpful to define $\chi_n(x)=X_n(x)/\sqrt{x(1-x)}$. %, with $X_n(x)$ being the function used in
%Refs.~\cite{DeTeramond:2021jnn,Ahmady:2021lsh,Ahmady:2021yzh,Chabysheva:2012fe}.  

One expects that  the correct QCD-potential is not a sum of two independent terms. In that case the space of product wave functions forms a useful, complete, relativistic basis, such as advocated in~\cite{Chabysheva:2012fe} and implemented in~\cite{Li:2015zda}. Nevertheless, Refs.~\cite{DeTeramond:2021jnn,Li:2021jqb,Ahmady:2021lsh,Ahmady:2021yzh} compare the resulting values of $M_h^2$ with measured mesonic spectra.

Next, we consider the equation of the form $H_\parallel \chi_n = M_n^2 \chi_n$, with $H_\parallel$ Hermitian. This is the form of the wave equation used by both 't Hooft and Callen, Coote \& Gross~\cite{Callan:1975ps} (and many others).
Note that the $V_{||}$ of Refs.~\cite{Li:2021jqb}  and ~\cite{tHooft:1974pnl} are not diagonal in $x$. This is because confining potentials must have an  explicit dependence on the coordinate-space variable, $\tz$,  that is canonically conjugate to $x$.  We shall see that none of the models of $V_\parallel$ in current use is of the form $V_{||}(x)$.

 With  the fundamental longitudinal wave equation  the usual orthonormal equation applies 
\begin{align}
    \int\,dx \chi^{*}_n(x)\chi_m(x) = \delta_{nm},
    \label{Norm}
\end{align}
as explicitly stated in Ref.~\cite{ Callan:1975ps}.
Next, consider the matrix element. 
\begin{align}
    \langle n |H|m\rangle &= \int dx\,dy\, \chi^*_n(x) H(x,y) \chi_m(y).%= \delta_{nm}M_n^2,
\end{align}
The  Hamiltonian can be expressed as
$%\begin{align*}
    H(x,y) = \frac{m^2}{x(1-x)} \delta(x-y) + V_{||}(x,y),\,
$%\end{align*}
 so that: 
\begin{align}
    \langle n|H|m\rangle &= \int dx \frac{m^2}{x(1-x)} \chi_n^{*}(x)\chi_m(x) \nonumber\\&+ \int \,dx\,dy\chi^*_n(x)V_{||}(x,y) \chi_m(y).
\end{align}

The longitudinal potential $V_\para$ must be chosen to determine the function $\chi_n(x)$. There are two choices in the literature. The first two   (LV) \cite{Li:2021jqb} and the 't Hooft model (tH) ~\cite{tHooft:1974pnl}, are given by
\bea
V_{\rm LV}(x)\chi_n(x)= -\s^2 \partial_x\,x(1-x)\partial_x\chi_n(x)\nonumber\\
(V_{\rm tH}\,\chi_n)(x)={g^2\over \pi} P\int_0^1dy  {\chi_n(x)-\chi_n(y)\over (x-y)^2},
\label{V}
\eea
with the principal value is   defined~\cite{PhysRevD.19.3024} as
$ P {f(x,y)\over (x-y)^2}\equiv {1\over 2}[ {f(x,y)\over (x-y +i \epsilon)^2}+{f(x,y)\over (x-y -i \epsilon)^2}]
$ in the limit $\e\to0$.
The function $V_{\rm LV}(x)$ is used in \cite{Li:2021jqb}, and has the advantage that exact solutions are available in terms of Jacobi polynomials. The  't~Hooft 
model~\cite{tHooft:1974pnl}, obtained in the large-$N$ limit of two-dimensional
QCD, is used in  \cite{Ahmady:2021lsh,Ahmady:2021yzh,Chabysheva:2012fe}. This 
 is the natural choice for a confining
potential in one spatial dimension.  When QCD is quantized in
light-cone gauge, such a potential appears automatically as
an instantaneous Coulomb-like interaction, $V_{\rm tH}(\tz)=g^2|\tz|e^{-\e|\tz|}$, between quark currents~\cite{PhysRevD.19.3024}, with $\tz$ as the longitudinal position operator~\cite{Miller:2019ysh} that is canonically conjugate to the longitudinal momentum variable $x$.
Taking the Fourier transform of $V_{\rm tH}(\tz)$ using the transformation $\la x|\tz\ra=e^{-i x \tz}/\sqrt{2\pi}$  and including  effects of the quark self-energy via the term $\chi(x)$ term   of the principle-value integral
 leads to the expression appearing in \eq{V}. Note that in the 't~Hooft model, the masses $m_{1,2}$ are explicitly current quark masses. 
 The 't Hooft model was extensively studied during the 1970's; see the review~\cite{Ellis:1977mw}.  
  
  The third approach \cite{DeTeramond:2021jnn} is to assert that $\chi(x)$ is a Gaussian in the invariant mass-squared: $\chi(x)={\cal N}\exp{[-1/(2\k^2)(-m_1^2/x+m_2^2/(1-x))]}$, where $\cal N$ is a normalization constant~\cite{Brodsky:2014yha}. This model is termed the invariant mass wave function (IMWF).
 
% Our analysis begins by explaining our use of $\chi(x)$ instead of $X(x)$.
%It has been typical to express the value of $M_\para^2$    as   
%\begin{align}
   % M_\para^2 &= \int \frac{ d x}{x(1-x)} \left[\frac{|X(x)|^2 m^2}{x(1-x)} + X^{*}(x) V_{||}(x) X(x)\right].
    %\end{align}%
%Then with $\chi(x) \equiv X(x)/\sqrt{x(1-x)}$  we obtain 
%\begin{align}
    %M_\para^2 &= \int\,\,dx \chi(x)\left[ \frac{m^2}{x(1 - x)} + V_{||}(x)\right]\chi(x)
%\end{align}
%using \eq{norm}.
%Thus, there seems to be a choice of norm conditions. %But there is nothing wrong with using the one for $\chi$.
% For example, Chabysheva \& Hiller~\cite{Chabysheva:2012fe} use the normalization for $X(x)$ in the main body of their paper, but the one for  $\chi(x)$ in the Appendix    displaying  their  numerical technique.

To obtain an expression for the Hamiltonian using the $X$ normalization: define
$X_n(x) \equiv \sqrt{x(1-x)}\chi_n(x)$ so that
\begin{align}
    \langle n|H|m\rangle &= \int \frac{dx}{x(1-x)}  {\tilde X}_n(x)\frac{m^2}{x(1-x)}X_m(x) +\langle n|V_\para|m\ra, 
    %\\& + \int \,dx\,dy\frac{\tilde X^*_{n}(x)}{\sqrt{x(1-x)}}V_{||}(x,y) \frac{X_{m}(y)}{\sqrt{y(1-y)}}.
\label{Hnm}\end{align}
with
\begin{align}
    \langle n|V_{||}|m\rangle &= \int \frac{dx\,dy}{x(1-x)}{\tilde X}_n(x) {\sqrt{x(1-x)}\over \sqrt{y(1-y)}}V(x,y) X_m(y).
\end{align}
The  effective potential, ${\sqrt{x(1-x)}\over \sqrt{y(1-y)}}V(x,y)$  appearing  between the two wave functions is not Hermitian. The function $\tilde X_n$ are obtained using the effective potential ${\sqrt{y(1-y)}\over \sqrt{x(1-x)}}V(x,y)$.

The consequence of this lack of hermiticity is that, for example,  the resulting wave equation for the  t' Hooft model would be  of the  form:
\begin{align}
&M_n^2 X_n(x)=\nonumber\\&    \frac{m^2}{x(1-x)}X_n(x) - \frac{g^2}{\pi} \mathcal{P}\int\,dy \frac{(X_n(x)-\frac{\sqrt{x(1-x)}}{\sqrt{y(1-y)}} X_n(y))}{(x-y)^2} .
\end{align}
This equation was solved in Ref.~\cite{Chabysheva:2012fe}.
The non-Hermitian nature of the effective potential seems to  represent a difficulty in extending   light-front holographic QCD in the conformal limit  to include non-zero quark masses and excitation of longitudinal modes.
%Nevertheless, the product wave functions of the form $\phi\,X_n$ can  be and have been  used as a basis set for  diagonalizing %$H_{\text{QCD}}$. 
Instead, the normalization given in \eq{Norm} is used for the rest of the paper.

We now begin to discuss the differences between the two  potentials of \eq{V}. 
In the chiral limit, the ground state wave function of both~\cite{Li:2021jqb,Callan:1975ps} can be seen from the differential equation for $M_\para^2$ to be simply ( dropping the subscript) $\chi(x)=1$, with an eigenvalue $M_\para^2=0$ . Ref.~\cite{DeTeramond:2021jnn} also have $\chi(x)=1$ in the chiral limit.

 The LFHQCD formalism~\cite{Brodsky:2014yha} achieved   excellent reproduction of the hadronic spectrum.  Using $\chi(x)=1$ along with $M_\para^2=0$ preserves that ground-state spectrum.  %Ref.~\cite{Chabysheva:2012fe} used $\chi(x)\propto \sqrt{x(1-x)}$ following the normalization convention of~\cite{Brodsky:2014yha}, which is not that of  solutions of the 't Hooft equation.

The useful identity~\cite{Callan:1975ps}:
\bea 
M_\para^2\int_0^1\,dx\,\chi(x)=\int_0^1\,dx\,\chi(x)\left[ \frac{m_1^2}{x}+\frac{m_2^2}{1-x} \right]
\label{tHm}\eea
holds in both the Li-Vary and 't Hooft models. The need to have finite results for both sides of this equation signifies that  $\chi(x)$ vanishes at the end-points, $x=0,1$.

 Given the similarities between the Li-Vary and 't Hooft models, it is natural to initially expect that  
any differences between the models might be thought to  be small.  This is far from the case. Using coordinate space helps in 
 understanding that there indeed are significant  differences between the potentials.   
This is done by evaluating $V_{\rm LV}(x)$ in coordinate space.
\bea& \la\tz| V_{\rm LV}(x)|\tz'\ra=\s^2 \tz\tz'\int_0^1 {dx\over 2\pi}e^{i x(\tz'-\tz)}x(1-x)\nonumber\\&
= \frac{\s^2}{2\pi}\tz\tz' e^{i\,{(\tz'-\tz)\over 2}}{j_1({\tz'-\tz\over 2})\over \tz'-\tz},
\eea
where $j_1$ is a spherical Bessel function. The diagonal elements of this operator are given by $\s^2\tz^2/6$ which is in stark contrast to the linear behavior of the potential in the 't Hooft model. Moreover, there is a significant coordinate-space non-locality in $V_{\rm LV}(x)$  that does not occur for $V_{\rm tH}$. Another difference is that the potential $V_{\rm LV}(x)$ is   complex although Hermitian.

The most salient  difference between the two potentials is in the high-energy spectrum:
$ M^2_\para({\rm tH)}\approx g^2 k,$~\cite{Callan:1975ps}
where $k$ is an integer. The high energy spectrum of the 't Hooft model is much more compact than that of the Li-Vary model. For large values of $k$, $\chi_{\rm tH}(x)=\sqrt{2}\sin(\pi k x +\delta(k,x))$. The phase shift $\delta(k,x)$ is given in Ref.~\cite{PhysRevD.19.3024}.

The spectrum corresponding to the  Brodsky-de Teramond  IMWF model~\cite{DeTeramond:2021jnn}    has not been provided before. The lore is that this model does not come from a wave equation. Here we develop a wave equation of the Sturm-Liouville form that does yield the IMWF and discuss the properties of the solutions. 
The first step is  asserting that  the IMWF is the solution  of some  wave equation.  The presentation is simplified by using $m_{1,2}=m$ and also by employing the variable $y>0$, with $y^2=1/(x(1-x)).$ Then using $\phi(y)=e^{-m^2y^2\over2\k^2}$,
consider the differential equation:
\bea-{d\over dy } f(y) {d\phi(y)\over dy}+m^2y^2\phi(y)=M^2\phi(y),\label{BdT}\eea
with $f(y)$ to be determined. The term $m^2y^2$ is the familiar kinetic energy term.
Using the expression for $\phi(y)$ and equating the left- and right-sides of the equation leads to the results $f=\k^4/m^2$ and $M^2=\k^2$. A peculiar feature is that the 
ground-state mass is independent of  $m$.
The wave function that leads to the ground state IMWF is then given by
\bea -{\k^4\over m^2}\phi''+m^2 y^2\phi=M^2\phi,\label{BdT1}\eea
which is of a familiar harmonic oscillator form.
%where both $m^2$ and $M^2$ are expressed in units of $\k^2$, an interaction strength. This is the familiar harmonic oscillator problem, with ground state solution
%$\phi(y)=N e^{-{m^2\over 2}y^2}$.  Make the substitution: $y^2 \to 1/(x(1-x))$, so that $\phi(y)\to  N \exp(-m^2/(2x(1-x))$, the desired form. Then using 
%\bea
This can be converted to an equation  in the variable $x$ by using
 ${dx\over dy}={1\over x-1/2}x^{3/2}(1-x)^{3/2}\equiv g(x), \,$ so that % \eea 
\eq{BdT1} becomes 
\bea -g(x){\k^4\over m^2}{d\over dx} g(x){d\phi(x) \over dx}+{m^2\phi(x)\over  x(1-x)} = {M^2} \phi(x),\eea
which is of the Sturm-Liouville form for $0<x<1/2$ and $1/2<x<1$. The ground state energy is  $\k^2$. Each excitation  in energy increases $M^2$ by 
 $2\k^2$.  The feature that the spectrum does not depend on quark masses does not seem reasonable and we therefore  discard \eq{BdT1}. 
 
The next step is to compare two remaining approaches~\cite{Li:2021jqb,tHooft:1974pnl} in the limit that the quark masses are non-zero, but small compared to the strength parameter of the model. For simplicity we remain with  the case that $m_{1,2}=m$.
The Li-Vary result is that 
\bea M_\para^2({\rm LV}) =2\s m +4m^2.\eea
Their model uses the parameters  $m=15 $ MeV, and $\s=620$ MeV.
't Hooft found that 
the ground-state wave function $\chi(x)$ is well approximated by
the form $\chi_0(x)=x^{\beta}(1-x)^{\beta}$.
Analysis of the  behavior for the solution of the wave equation
%(\ref{eq:longitudinaleqn}) 
for $x\to0\,\rm or\,1$ shows that $\beta$ should satisfy the
transcendental equation
$
\frac{m^2 \pi}{g^2}-1+\pi\beta \cot\pi\beta=0.
$ %\ee
For small values of $m$, the quantity $\b \pi$ must also be small so that $\b=\sqrt{3\over \pi}{m\over g}$.
Using  $\chi_0$ on both sides of  \eq{tHm} yields that
$  M_\para^2=% {\int dx {m^2\over x(1-x)}X(x)\over \int dx X(x)}=\frac{\Gamma (\beta )^2 \Gamma (2 \beta +2)}{\Gamma (2 \beta ) \Gamma (\beta +1)^2}=
m^2(\frac{2}{\beta }+4),$ or equivalently
 \bea
  M_\para^2({\rm tH})=2\sqrt{\pi\over 3}g      m+4    m^2.
  \eea

Using  the average value of the $u$ and $d$ quark masses, $m=3.5$ MeV~\cite{Zyla:2020zbs} with $M_L^2=140$ MeV,  tells us that      
  $g=2700$  MeV, and  $\b=0.00126.$                           
  Thus, these two  models contain a  (1+1)-dimensional version  of the  Gell-Mann-Oakes-Renner~\cite{Gell-Mann:1968hlm} relation in which the squared mass of the ground state is proportional to the current quark mass.        
  
  \begin{figure}[h]
\vspace{0.15in}
\begin{center}
   \begin{subfigure}[b]{0.535\textwidth}
   \hspace{-0.2in}
   \includegraphics[scale = 0.4,trim = {0cm 0cm 0cm 0cm}, clip]{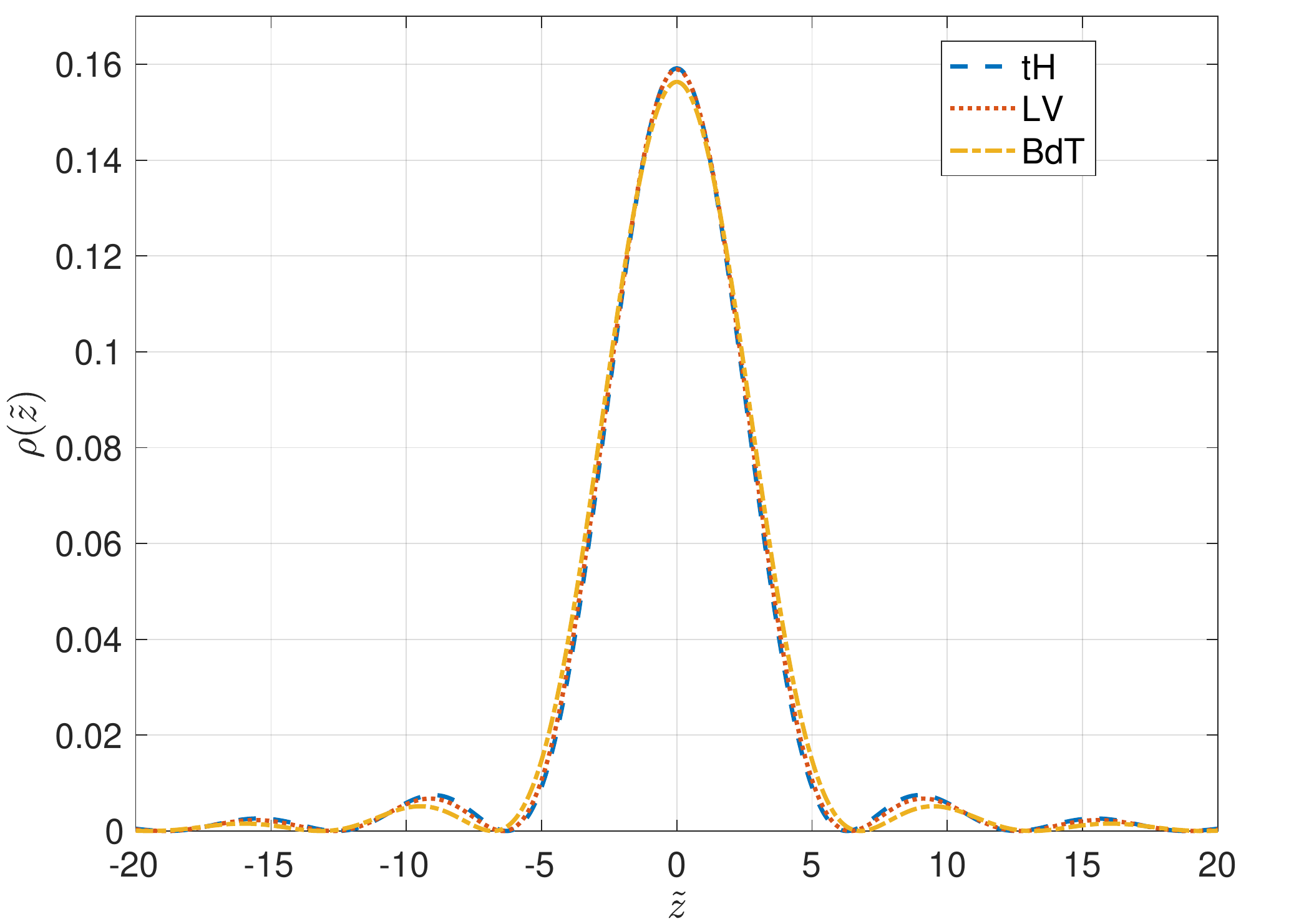}
   \caption{}
   \label{fig:Ng1} 
\end{subfigure}
\end{center}
\begin{subfigure}[b]{0.535\textwidth}
   \includegraphics[scale = 0.4,trim = {0cm 0cm 0cm 0cm}, clip]{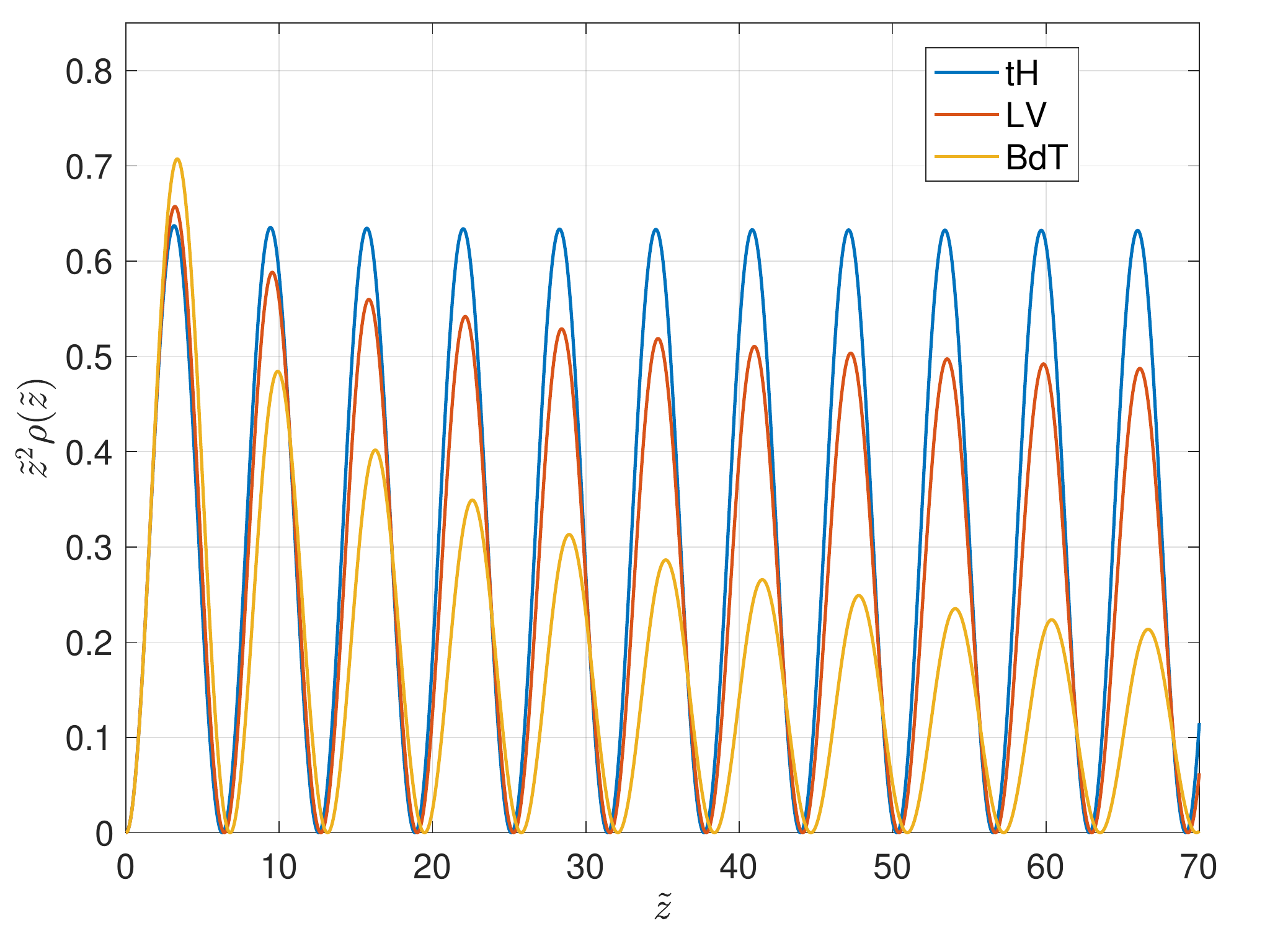}
   \caption{}
   \label{fig:Ng2}
\end{subfigure}
\caption{Plots of the resulting density, $\rho(\tilde{z})$, and $\tilde{z}^2\rho(\tilde{z})$ are given for all three models. The discrepancy between the models in $\rho(\tilde{z})$ is small but finite, and the $1/\tilde{z}^2$ asymptotic behavior is confirmed.}
\end{figure}
It is worthwhile to note that the use of  current-quark masses in the 't Hooft model causes the  two different models to obtain very different masses of the first excited state.
  Ref~\cite{Li:2021jqb} obtains  excitation energies of approximately 1 GeV and associates these values with excited states of the pion. In the present work, using modern values of the current quark masse ~\cite{Zyla:2020zbs}  gives the lowest excited state a mass on the order of $g$, or about 3 GeV. This is high enough into the continuum of states with large widths to be unobservable. Thus, the  version of the 't Hooft model used here preserves the  spectra produced by LFHQCD.

The confining aspects of the 't Hooft model have been well-studied long ago~\cite{tHooft:1974pnl,Callan:1975ps}, using a momentum-space ($x$) dependence approach based on studying the cancellation of infrared singularities. Another, possibly more intuitive approach, may be obtaining by examining coordinate-space wave functions that depend on the canonically conjugate spatial variable, $\tz$. An intuitive way to think about this variable is that it is the separation between the quark and anti-quark in the direction of motion of a pion moving with high momentum.

Coordinate-space wave functions  are obtained using the transformation
 \bea\chi(\tz)\equiv \int_0^1{dx\over \sqrt{2\pi}} e^{i x \tz}\chi(x).
%= \int {dx\over \sqrt{\pi}} e^{i x \tz}\sin{n\pi x}\\={n\pi\over \sqrt{\pi}} {-1+(-1)^n\,e^{iz}\over z^2-n^2\pi^2}\\
%\phi_n(n\pi)={i\over 2}
\eea
%One could use the replacement $\chi(x) \to X(x)\over \sqrt{x(1-x)}}$, which makes no change 
It is useful to examine the density $\r(\tz)\equiv |\chi(\tz)|^2$, a real-valued quantity,  for the three models are shown in Fig.~1(a).
%BdT $m=28$ MeV, ${\k}$ =0.523 GeV. Li-Vary $m=15 $ MeV, $\s=620$ MeV; tH $g=2713.7 MeV$ , $\b=0.00126.$
These results seem very similar  because of the relatively small quark masses of the three models. The densities seem to go to 0 for large absolute values of $\tz.$ They do,  but there in an interesting feature seen by plotting $\tz^2\r(\tz)$ in Fig.~1(b): the densities fall as $1/\tz^2$.   

This behavior  may be understood by making a asymptotic expansion, obtained by using $e^{i\tz x}=1/(i \tz) \partial_xe^{i\tz x}$ and the feature that
$\chi(x)$ vanishes at the end points. Then
 \bea\lim_{|\tz|\to \infty}\chi(\tz)={i\over  \tz} \int_0^1 {dx\over \sqrt{2\pi}}  e^{i x \tz} \partial_x\chi(x).\eea
 Squaring this quantity leads to the stated $1/\tz^2$ dependence.

 The similarity of the behaviors for small masses suggests that  the chiral limit should be examined. In this case, $\chi(x)=1$ for all three models. Then a simple  closed-form expression for $\chi(\tz)$ can be obtained. The resulting density, $\r_\chi(\tz)$ is given by
 \bea \r_\chi(\tz)={2\over \pi}  {\sin^2{\tz/2}\over \tz^2},\eea
 an expression that accounts explicitly for the oscillatory behavior as well as the $1/\tz^2$ asymptotic behavior. One may also examine the spatial extent of the pion wave function by  considering the mean-square value of $\tz^2$, that is given by the   ground-state expectation value:
 \bea \la\chi|\tz^2|\chi\ra ={2\over \pi}\int_{-\infty}^\infty\,dz\,\sin^2(\tz/2)=\infty.\eea
In the chiral limit the pion has an infinite spatial extent, true for all three models.

It is necessary to see how or if this infinite size conflicts with current understanding. First note that the elastic pion form factor of LFHQCD has already  been 
computed using $\chi(x)=1$; see {\it e.g.} Ref.~\cite{deTeramond:2018ecg}. The infinite extent is buried within the integrals needed to compute the  elastic form factor, a consequence the  ability to   probe with only  transverse momentum transfers; see {\it e.g.} the review~\cite{Miller:2010nz}.

The relevance of the infinite longitudinal extent can be understood  in analogy with the familiar Ioffe-time argument \cite{Gribov:1965hf,Ioffe:1969kf,Kovchegov:2012mbw,Braun:2007wv} for deep inelastic scattering at small values of Bjorken $x$. The idea is that  an incident virtual photon fluctuates into a $q\bar q$ pair. The energy difference, $ \D E$,
 between the two states is very small, so that the fluctuation lives for a long time $\D t>1/\D E$.  For virtual photons of high energy, the extent of the fluctuation is $c\D t$ is very large. Now consider a high-momentum pion incident on a target. The value of $\D E $ goes to 0 if both the pion and its constituents are massless. Thus the infinite extent of a mass-less pion is part of standard lore.

We summarize. The similarities and differences between the three models of Refs.~~\cite{DeTeramond:2021jnn,Li:2021jqb,tHooft:1974pnl} are exhibited. Very significant differences in the excitation spectra, at both low and high energies,  are obtained, even though all of  the ground-state wave functions are the  same in the chiral limit. If the 't Hooft model is used along with current values of light-quark masses, the original spectrum calculations reviewed in Ref.~\cite{Brodsky:2014yha} are preserved. Finally, the pion is shown to have an infinite extent in the longitudinal direction.\\

 \acknowledgments
This work was supported by the U.S. Department of Energy Office of Science, Office of Nuclear Physics under Award No. DE-FG02-97ER-41014.
We thank J. Hiller, Y.~Li, J. P. Vary, R. Sandapen, G.F. de~T\'eramond and S.J. Brodsky for helpful
comments.

\appendix

 \bibliography{millerOCT12}
\end{document}